%% file: _main.tex
\documentclass[11pt]{article}
\usepackage[final]{acl}
\usepackage[a-1b]{pdfx}
\input{0_use_packages}
\input{0_macro}

\title{Discovering Biases in Information Retrieval Models Using Relevance Thesaurus as Global Explanation}
\author{Youngwoo Kim, Razieh Rahimi, \and James Allan \\
        University of Massachusetts Amherst
 \\ \texttt{\{youngwookim, rahimi, allan\}@cs.umass.edu}}

\begin{document}
\maketitle

\begin{abstract}
Most efforts in interpreting neural relevance models have focused on local explanations, which explain the relevance of a document to a query but are not useful in predicting the model's behavior on unseen query-document pairs. We propose a novel method to globally explain neural relevance models by constructing a ``relevance thesaurus'' containing semantically relevant query and document term pairs. This thesaurus is used to augment lexical matching models such as BM25 to approximate the neural model's predictions. Our method involves training a neural relevance model to score the relevance of partial query and document segments, which is then used to identify relevant terms across the vocabulary space. We evaluate the obtained thesaurus explanation based on ranking effectiveness and fidelity to the target neural ranking model. Notably, our thesaurus reveals the existence of brand name bias in ranking models, demonstrating one advantage of our explanation method.
\footnote{Code and results are available at \url{https://github.com/youngwoo-umass/RelevanceThesaurus}}
\end{abstract}

\input{1_intro}
\input{2_related}
\input{3_0_method}

\input{4_experiments}
\input{5_conclusion}

\section*{Acknowledgement}
This work was supported in part by the Center for Intelligent Information Retrieval and in part by NSF grant \#2106282. Any opinions, findings and conclusions or recommendations expressed in this material are those of the authors and do not necessarily reflect those of the sponsor.

\input{8_limitations}
\bibliography{_main}
\newpage

\input{9_0_appendix}

\end{document}

%% file: 0_use_packages.tex
\usepackage{savesym}
\savesymbol{dddot}
\savesymbol{ddddot}
\usepackage{amsmath}
\usepackage{accents}
\usepackage{amsfonts}
\usepackage{array}
\usepackage{booktabs}
\usepackage{epsfig}
\usepackage{graphicx}
\usepackage{latexsym}
\usepackage{longtable}
\usepackage{makecell} 
\usepackage{multirow}
\usepackage{pgfplots}[style=base]
\usepackage{soul}
\usepackage{subcaption}
\usepackage{tabularx}
\usepackage{tabu}
\usepackage{threeparttable}
\usepackage{times}
\usepackage{tikz}
\usepackage{xcolor,colortbl}
\usepackage[utf8]{inputenc}
\usepackage{bbold}
\usepackage{xspace}
\usepackage{supertabular}
\usepackage{multicol}
\usepackage{minitoc}
\usepackage{subcaption}
\usepackage{tcolorbox}

\pgfplotsset{compat=1.18}
\usetikzlibrary{fit,tikzmark}
\usetikzlibrary{positioning}

%% file: 0_macro.tex
\newcommand{\topic}[1]{{}}

\newcommand{\hide}[1]{{}}
\newcommand{\ebegin}[1]{{\bf \color{purple} $>>>>>$ Beginning of edits $>>>>>$ \\} \color{violet}}
\newcommand{\eend}[1]{{\bf \color{purple} $<<<<<$ End of edits $<<<<<$  \\} \color{black}}

\renewcommand{\cite}[1]{{\citep{#1}}}

\makeatletter
\if@twocolumn
  
\else
  
\fi
\makeatother

\newcommand{\modelname}{PaRM\xspace}

%% file: 1_intro.tex
\section{Introduction}
Transformer-based information retrieval (IR) models~\cite{dai2019deeper, macavaney2019cedr} that are trained on large datasets like MS MARCO~\cite{msmarco} are very effective in predicting relevance between a query and document. 
Contextual representations in these models enable semantic matching, such as matching the query term ``car'' with the document term ``vehicle''.
However, it is challenging for researchers to predict the potential failures of a model, such as when it matches a query term to non-relevant document terms. 

Another potential risk associated with neural retrieval models is an unintended bias toward certain entities or groups~\cite{may2019measuring}.
While it is appropriate for a model to associate the query term ``car'' with various car brand names (e.g., Ford), the model should not exhibit a strong preference for a particular brand, leading to the model favoring that brand over another when all other factors are identical.
For the safe deployment of information retrieval~(IR) models in real-world scenarios, detailed global understanding of model behavior are essential, such as providing which lexical expressions are considered relevant by the models.  
 \input{table/ex_table}

To address these challenges and mitigate potential risks, post-hoc explanation methods for black-box machine learning models can be employed. 
Most explanations for IR model explanations are local explanations,  focusing on individual model predictions, such as a specific query-document pair~\cite{kim2022alignment} or a ranked list for a query~\cite{LIRME, llordes2023explain}. These explanations indicate which terms in the documents contribute to its relevance to the query. However, local explanations have two major limitations that hinder their ability to infer cases where the model may exhibit unexpected behavior.

First, the explanations are limited to the terms observed in the given query and document, and biases may exist in queries or documents that were not evaluated or inspected with the explanations. Second,  attribution to document terms by explainers may be highly dependent on the contexts of those terms, therefore it is unclear whether the attributed document terms in other contexts would match the query. 

To overcome these limitations, we propose building a global explanation~\cite{guidotti2018survey} that provides lexical insights about query-document terms that are matched by the model in all contexts. We can describe a model's behavior in a compact and interpretable structure that is not limited to a specific instance. 

Our proposed global explanation focuses on identifying relevant pairs of query and document terms that can effectively explain the matching behavior of neural retrieval models. We refer to this format of explanation as a \textit{relevance thesaurus}, with examples illustrated in Table~\ref{ex:table}. The table indicates that, if a query contains term ``injury,'' then it is likely for the model to match the query term with document terms ``injure,'' or ``wound,'' with the former being the more likely. This allows researchers to anticipate which terms, when present in a document, would lead the model to predict higher relevance for that document, without requiring additional context from the document.


Constructing a relevance thesaurus is challenging due to the large number of potential term pairs.  Many local~\cite{should} and global explanation~\cite{han2020explaining} methods build a candidate set of features from data and adjust their scores based on the target model's outputs. However, this approach becomes infeasible when the number of features reaches to billions, as in our study. To overcome this challenge, we propose a novel approach that distill the knowledge of the target model into an intermediate neural model, PaRM (Partial Relevance Model), which is then used to infer important features. 

PaRM is designed to predict a score for a term pair, which is then used to predict the score for the corresponding query-document pair. By training PaRM with knowledge distillation from the target neural model to be explained, we ensure that the generated relevance thesaurus faithfully explains the target model's behavior.

Rather than assessing the accuracy of each term pair in the relevance thesaurus individually, the thesaurus is extrinsically evaluated by integrating it into lexical matching models (BM25~\cite{robertson2009probabilistic} and QL~\cite{ponte1998language}), adding interpretable semantic matching to them.
The resulting retrieval methods are evaluated based on retrieval effectiveness and fidelity to the target neural retrieval models.
The results on multiple datasets show the effectiveness of the acquired relevance thesaurus.

To demonstrate the advantages of our relevance thesaurus, we introduce three unexpected findings about the behavior of neural retrieval models trained on MS MARCO, obtained from our analysis of the relevance thesaurus: 
(1) the \textit{car-brand bias}, which suggests that models exhibit biases towards certain car brands; (2) the \textit{temporal bias}, which indicates that models consider distant future or past years to be more strongly associated with the query term ``when'' compared to the current year;
(3) the \textit{postfix-a} finding, which reveals that models treat the character ``a'' appended to a term as equivalent to a quotation mark due to encoding errors.

Experiments using multiple state-of-the-art neural information retrieval models demonstrate that these behaviors are not limited to the cross-encoder ranker which is used to distill the relevance thesaurus but are also the case in multiple other IR models, Splade~\cite{formal2021splade} and Contriever~\cite{izacard2021unsupervised}. This highlights the importance of global explanations for retrieval models.
\input{figures/outline}

%% file: table/ex_table.tex

\newcommand{\drawBox}[2]
{
  \begin{tikzpicture}[node distance=0mm]
  \tikzset{Nums/.style={font=\scriptsize, inner sep=0pt}}
  \tikzset{Word/.style={font=\small, minimum height=13pt}}
  \node[Word] (word){#1};
  \node[below=of word, Nums]{#2};
  \end{tikzpicture} 
}
\newcommand{\hlTikZ}[1]{%
  \tikz[baseline=(X.base)]{
    \node[rectangle, fill=red!30, inner sep=1pt] (X) {#1};
  }%
}

\begin{table}[t]
\centering
\small
\begin{tabular}{
>{\centering\arraybackslash}m{16mm}|
>{\centering\arraybackslash}m{10mm} 
>{\centering\arraybackslash}m{9mm} 
>{\centering\arraybackslash}m{9mm} 
c}
\toprule
 \textbf{Query term} & \multicolumn{4}{c}{\textbf{Document term}} \\ \hline
injury     & \drawBox{injure}{0.26}   & \drawBox{wound}{0.24}& \drawBox{torn}{0.19} & ... \\ \hline
when       & \drawBox{24th}{0.33}     & \drawBox{2010}{\hlTikZ{0.11}} & \drawBox{2015}{\hlTikZ{0.01}} & ...\\ \hline
car        & \drawBox{vehicle}{0.68} & \drawBox{ford}{\hlTikZ{0.38}} & \drawBox{honda}{\hlTikZ{0.28}}& ...  \\ \hline
cud        & \drawBox{cudâ}{0.50}     & \drawBox{\hlTikZ{cuda}}{\hlTikZ{0.50}} & ... & \\ \bottomrule
\end{tabular}
\caption{Example entries from our relevance thesaurus. The numbers indicate the degree of relevance. Unexpected behaviors found by our method are highlighted. }
\label{ex:table}
\end{table}

%% file: figures/outline.tex
\begin{figure*}
    \centering
    \includegraphics[width=\textwidth]{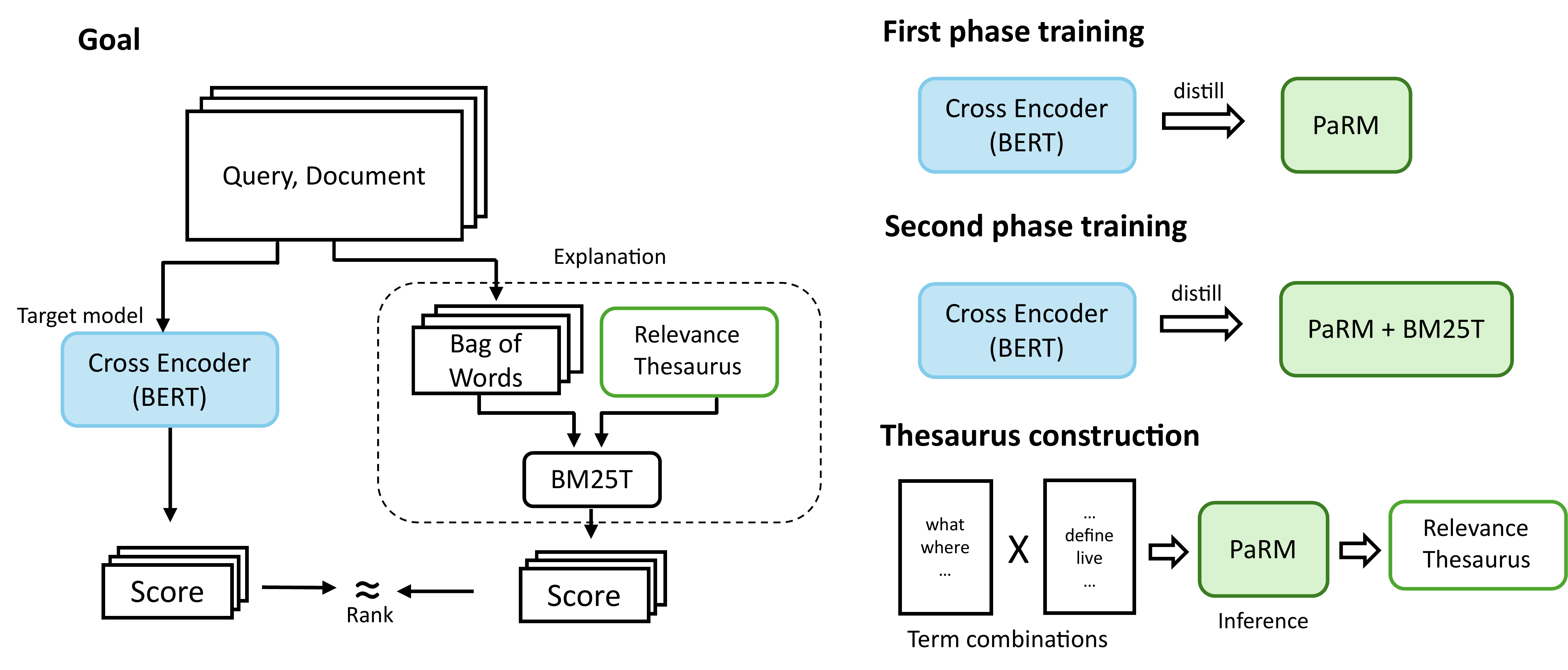}
    \caption{Our goal is to build a relevance thesaurus that can approximate the cross-encoder model (left). The relevance thesaurus is expected to be generalizable to any queries. The figure on the right shows how the relevance thesaurus is constructed. The colored boxes are black-box models, and the white boxes are interpretable components.}
    \label{fig:overview}
\end{figure*}

%% file: 2_related.tex
\section{Related works}

\subsection{Global model explanations}
Large portions of works on global explanations are for classification tasks on tabular features~\cite{craven1995extracting, boz2002extracting, guidotti2018survey}.
They cannot be applied to the Transformer model for token sequences,  as tabular features are not defined.
Instead, global explanation works in the NLP domain target single text classification, by attributing output labels to some words or phrases~\cite{rajagopal2021selfexplain, han2020explaining}. 
However, these word-to-output label attributions are not applicable for explaining text pair models like IR models, because document terms' importance is highly dependent on queries.
It would make a more meaningful explanation if it indicate certain terms or phrases from the query are associated with specific terms or phrases that appear in the document.

\subsection{Explanations for neural IR}
Existing neural IR models~\cite{khattab2020colbert, gao2021coil, formal2021splade, nogueira2019doc2query, kim2021query} encode entire queries and/or documents with a single Transformer network and are not capable of encoding parts of the query/document in the absence of the remaining context. Applying perturbation-based explanation approaches~\cite{kim2020explaining, should} on these models can be problematic because removing tokens from a query could lead to a larger change of the meaning in IR tasks than the other NLP tasks. For example, a document that is relevant to the query ``ACL location'' is not relevant to the query ``location'', as the relevant document for this query needs to describe what the expression ``locations'' means rather than mentioning any location. 

Existing IR models explanations~\cite{LIRME, llordes2023explain, lyu2023listwise, pandian2024evaluating, naseri2021ceqe} mostly work in the query-level, and output terms for one query cannot be used to infer the model's behaviors in other queries. \citet{chowdhury2023rank} target explaining categorical features in learning-to-rank IR models, while we target sequence processing Transformer models. 



To enhance lexical models with recent advances, \citet{boytsov2021exploring} proposed fine-tuning BERT~\cite{devlin-etal-2019-bert} for the translation language model~\cite{berger1999information}. This approach, however, is limited to the semantic matches between terms in BERT's subword vocabulary and does not extend to terms formed from multiple subwords. Moreover, the work lacks analysis or evaluation regarding the explanation perspectives and does not provide qualitative insights from the outcomes.

\subsection{Interpretable NLP models}
Our proposed model architecture is motivated by the series of the work on the natural language inference task~\cite{wu2021weakly, stacey2022logical, condnli}. Specifically, we adopted the idea of partitioning a sentence into two segments from the work by \citet{condnli}. 

%% file: 3_0_method.tex
\section{Relevance thesaurus building}
\label{sec:rel_term_metric}
We define the global  explanation of an information retrieval model, based on the definition by \citet{guidotti2018survey}, as follows:

\textbf{Model explanation problem.} Given a black-box relevance predictor $S_b$ that takes a query $q$ and a document $d$ as inputs and predicts a relevance score $y \in Y$, a global model explainer aims to find a human-interpretable explanation $E$ and an explanation logic $\epsilon$. The explanation logic $\epsilon$ is a function that converts the explanation $E$ into a global predictor $S_e$. The global predictor $S_e$ predicts a score $\hat{y}$ for $(q, d)$, which approximates the prediction  $y$.

As a black-box predictor $S_b$ to be explained, we target the full cross-encoder (CE) document ranking model~\cite{dai2019deeper}, which takes the concatenated sequence $q; d$ into the Transformer encoder to predict the relevance.
As a format for an explanation $E$, we use a relevance thesaurus, which is a set of triplets ${(qt, dt, s)}$, where $qt$ is a query term, $dt$ is a document term, and $s$ is the score assigned to the term pair.

To build an interpretable predictor $S_e$, we incorporate the relevance thesaurus ($E$) into the BM25 scoring function to address vocabulary mismatches between queries and documents. The relevance thesaurus captures semantic relationships between terms that may not be explicitly present in the query or document. Incorporating the relevance thesaurus improves the retrieval performance of the sparse retrieval model BM25, 
and provides an interpretable explanation of the CE model's behavior.


Many local and global explanation methods explicitly build a candidate set of features (e.g., terms) from data. 
To determine the appropriateness of each candidate feature as an explanation, these methods initially assign scores to the features. Then, the scores are adjusted based on the observed  behaviors of the model.
This strategy maintains explicit feature candidates and their scores during the optimization process.
This can be challenging as the number of features increases, especially in our explanation format where the number of term pairs can scale to billions.

To address this challenge, we propose  implicitly optimizing the explanation features using an intermediate neural model that scores features, namely term pairs. 

 As an intermediate neural model, we propose PaRM (Partial Relevance Model), which is designed to score relevance between partial segments of a query and document. 
 Unlike the original cross-encoder model and other relevance models that assign a score to an entire query and document, PaRM can predict meaningful scores for partial queries or documents.

This is important because the original CE model cannot accurately assess the contributions of individual tokens when they are isolated from their original contexts.
 For example, if the query is ``Who is Plato'' and the document is a single term ``Plato'', the original CE model would likely predict a score indicating non-relevance, as a document with a single term is unlikely to provide meaningful information. PaRM, on the other hand, predicts a score that indicates partial relevance, which can be combined with partial relevance scores for other terms to build the final relevance score for the query-document pair.
We are using the context independence assumption here.
This assumption is useful because it allows us to use any term pair predicted as relevant by PaRM to be globally indicative of relevance, which is not possible using local explanations.
\input{figures/parm_train}



PaRM is trained end-to-end by distilling predictions from the CE model. However, a challenge arises because PaRM is expected to predict a score for a query term and document term, while the available signal is only at the query-document level. 
To supervise term-pair level scores in PaRM using the CE model, alignments between query and document terms are required, but these are not directly available. The novelty of PaRM training lies in the ability to infer alignments in an unsupervised way.

In the first stage, we train PaRM to predict scores for two partial query-document segment pairs, using weak alignments, such as attention scores. After the first stage, we use the PaRM model to infer term-level alignments. 
In the second stage, we use the generated alignments by PaRM to further fine-tune the PaRM model to predict appropriately scaled scores for word-level relevance. These prediction are then  used to create the final explanations ($E$).

\subsection{PaRM first phase training }
The first phase PaRM predicts the relevance score $S_e$ for a given query-document pair by generating scores for two partial inputs, ($q_1, d_1$) and ($q_2, d_2$). These inputs are built by extracting a continuous span from the query $q$ to form $q_1$ and using the remaining tokens with a [MASK] token for $q_2$. 

The corresponding document segments, $d_1$ and $d_2$, are constructed by masking tokens from the document $d$ that have low attention scores for $q_1$ and $q_2$, as determined by the full cross-encoder (CE) ranker.

We randomly select how many tokens to be left in $d_i$, ranging from one to all tokens of $d$  (see \autoref{sec:appendix_sample} for details). In most cases, $d_i$ contains sufficient evidence to learn relevance while still allowing for a few extreme cases where either only a single query term or a single document term is present.

Scores for ($q_1$, $d_1$) and ($q_2$, $d_2$) are obtained by projecting BERT's CLS pooling representations.
\begin{equation}
    \text{PaRM}(q_i, d_i) = W \cdot BERT_{CLS}(q_i ; d_i) + b 
\end{equation}

The final score for the query and document pair is the sum of the scores from two partial views. 
\begin{equation}
    S_e(q, d) = \text{PaRM}(q_1, d_1) + \text{PaRM}(q_2, d_2)
\end{equation}

The combined score $S_e$ is trained from the scores ($S_b$) of the target black-box model (CE) using margin mean square error (MSE) loss~\cite{hofstatter2021efficiently} on relevant and non-relevant query-document pairs.
\begin{align}
\mathcal{L} = \text{MSE}(&S_e(q, d^+) - S_e(q, d^-),  \label{eq:mmse} \\
&S_b(q, d^+) - S_b(q, d^-)) \nonumber
\end{align}

Once PaRM is trained, we can use it to score an arbitrary query span or document span, including a single term. However, the scores are only trained for ranking and not calibrated to a specific range, which makes it hard to determine which term pairs have a sufficiently large score to be included in the relevance thesaurus.

\subsection{Fine-tuning PaRM with BM25} 
In the second phase, we fine-tune PaRM so that it scores the relevance of a query term $qt$ and a document term $dt$ on a scale from 0 to 1. Specifically, we consider the scenario of augmenting BM25 by handling vocabulary mismatch based on the scores from PaRM. 

We consider a query-document pair that any query term is missing in the document. We assume that the document term that has the highest PaRM score against the corresponding query term is most likely to be relevant to the query term if any term is relevant. We then use the output of PaRM to replace the term frequency (\autoref{fig:parm_stage2}). If the assumed pair is relevant, it will be more likely to appear in the relevant document and will be trained to score higher, and non-relevant ones will appear in the non-relevant document and be trained lower. 

For a pair of query $q$ and document $d$, if any query term does not have an exact match in the document, we randomly select one query term $qt$ to be trained. All document terms are scored against $qt$ using $\text{PaRM}(qt, dt)$ and the document term $dt$ with the highest score is paired with $qt$.
Note that terms are not from the BERT tokenizer, but are from the tokenizer developed for BM25. Thus, a single term can contain multiple BERT subwords. 

The training network is defined as follows. 
To ensure the output is between 0 to 1, we apply a sigmoid layer ($\sigma$) on top of the projected output. 
\begin{equation}
    \text{PaRM}(qt, dt) = \sigma(W \cdot \text{BERT}_{\text{CLS}}(qt ; dt) + b)~\label{term_sidmoid}
\end{equation}

In the original BM25 formula, the score for the query term $qt$ is determined by $qt$'s document frequency, $tf_{qt, d}$. We modify BM25 so that when a query term  does not appear in the document,  $tf_{qt, d}$ is replaced with the output of $\text{PaRM}(qt, dt)$.

\begin{align}
    f(qt,d) =
    \begin{cases}
    tf_{qt, d} & \text{if $qt \in d$} \\
    \text{PaRM}(qt, dt) & \text{if $qt \notin d$}
  \end{cases}~\label{term_case}
\end{align}
Note that $tf_{qt, d}$ can be large but PaRM is bounded above by 1, thus a non-exact match is never stronger than a single exact match.
The relevance score is computed based on the BM25 scoring function:

\begin{align}
    S_e(q,d)=\sum _{qt \in d}{
    \text{IDF}}(qt)\cdot {
    \frac {f(qt,d)\cdot (k_{1}+1)}{f(qt,d)+K}
    }  ~\label{eq:bm25t}
\end{align}
where $K$ is a function of document lengths, which is independent of $f(qt,d)$. 

PaRM is trained end-to-end from the pairwise hinge loss between a relevant pair $(q, d^+)$ and a non-relevant pair $(q, d^-)$:
\begin{equation}
\mathcal{L} = \max\left(0, 1 - S_e(q, d^+) + S_e(q, d^-)\right).~\label{term_pairwise}
\end{equation}

Note that we do not use knowledge distillation here, because the output scores scale of the BM25 scoring function is not easily adjustable and may not be possible to match the score margin of the neural ranking model.   

During the training phase,  equations  \ref{term_sidmoid} to \ref{term_pairwise} are implemented within a neural network framework, and the gradient to the loss $\mathcal{L}$ is back-propagated to train PaRM's parameters. Note that PaRM scores for selecting the highest scored $dt$ are pre-computed with the model after the first phase. 

After PaRM is fine-tuned, it can pre-compute the scores for potential $qt$ and $dt$ candidates. These candidate pairs and scores compose a relevance thesaurus. The acquired relevance thesaurus can be used to either inspect the model's behavior or used with the BM25 scoring function.

We name the modified BM25 function that addresses non-exact lexical matches based on the relevance thesaurus as BM25T (BM25 with Thesaurus).
For each query term $qt$, if $qt$ is found in the document $d$, relevance thesaurus is not used. If $qt$ is not found, the document term $dt \in d$ with the highest (pre-computed) $\text{PaRM}(qt, dt)$ score in the thesaurus is used to compute the score.

%% file: figures/parm_train.tex
\begin{figure}[t]
    \centering
    \begin{tikzpicture}[node distance=1pt]
    \tikzset{Token/.style={rectangle,minimum height=5mm, minimum width=3mm,}}
    \tikzset{QToken/.style={Token, fill=red!10, font=\small}}
    \tikzset{DToken/.style={Token, fill=blue!10, font=\small}}
    \tikzset{TextBox/.style={font=\small}}
    \tikzset{QLine/.style={->, thick, red!40}}
    \tikzset{DLine/.style={->, thick, blue!40}}
    \tikzset{Predictor/.style={rectangle,draw,minimum width=1.8cm,minimum height=1cm}}
    \tikzset{Prediction/.style={rectangle, fill=black!10, minimum height=3mm, minimum 
    width=3mm, font=\small}}
        
    \matrix[anchor=west, row sep=1mm] (left_tower) {
        \node[Prediction] (z1){$\rho_1$}; \\ [1mm]
        \node[Predictor] (ls1) {PaRM}; \\
        \node(q1)[QToken]  {$q_1$}; 
        \\
    };
    \node(d1)[DToken, minimum width=10mm, fill=blue!10, right=0.2em of q1] {$d_1$}; 
    
    \node[above=15mm of d1] (zs) {};
    \draw[<-] (z1.south) -| ++(0, -0.20);

    \node[right=of left_tower] (center) {};
    \matrix[anchor=west, row sep=1mm, right=of center] {
        \node[Prediction] (z2){$\rho_2$}; \\ [1mm]
        \node[Predictor] (ls2) {PaRM}; \\
        \node(q2)[QToken]  {$q_2$}; 
        \\
    };
    \node(d2)[DToken, minimum width=7mm, fill=blue!10, right=0.2em of q2] {$d_2$}; 
    \node[left=3em of q1] (q1left) {};

    \node[QToken, below=1em of q1left] (q1bot) {$q_1$};
    \node[TextBox, right=-2pt of q1bot] (qtext1) {who is [MASK]};   
    \node[QToken, below=of q1bot] (q2bot) {$q_2$};
    \node[TextBox, right=-2pt of q2bot] (qtext2) {[MASK] Plato};

    \node[DToken, right=of qtext1] (d1bot) {$d_1$};
    \node[TextBox, right=-2pt of d1bot] (dtext1) {[MASK] was a Greek philosopher.};
    \node[DToken, right=6pt of qtext2] (d2bot) {$d_2$};
    \node[TextBox, right=of d2bot] (dtext2) {Plato was a Greek [MASK] .}; 
    \node[rectangle,draw, right=7mm of z1] (combine) {\small{$Sum$}};
    \node[Prediction, above=2mm of combine] (y) {$y$};
    \draw[<-] (z2.south) -| ++(0, -0.20);

    \draw[->] (z1) -- (combine);
    \draw[->] (z2) -- (combine);
    \draw[->] (combine) -- (y);
    
    \end{tikzpicture}
      \caption{The first phase of training PaRM. The query ``who is Plato'' (red) is partitioned into $q_1$ and $q_2$. The document ``Plato was a Greek philosopher'' (blue) is masked to generate $d_1$ and $d_2$.}
    \label{fig:parm_train}
\end{figure}
\begin{figure}[t]
    \begin{tikzpicture}[node distance=1pt]
        \tikzset{Token/.style={rectangle,minimum height=6mm, minimum width=8mm,}}
        \tikzset{QToken/.style={Token, fill=red!10, font=\small}}
        \tikzset{DToken/.style={Token, fill=blue!10, font=\small, minimum width=15mm}}
        \tikzset{QCan/.style={draw=red!30, font=\small}}
        \tikzset{DCan/.style={draw=blue!30, font=\small, minimum width=15mm}}
        \tikzset{Predictor/.style={rectangle,draw,minimum width=1.8cm,minimum height=1cm}}
        \tikzset{Prediction/.style={rectangle, fill=black!10, minimum height=3mm, minimum width=3mm, font=\small}}    
        \tikzset{TextBox/.style={font=\small}}
        \tikzset{TFBox/.style={font=\small, rounded corners, anchor=west, draw, row sep=-1mm}}

        \matrix[TFBox] (qtf) {
            \node[QCan] {who: 1}; \\ 
            \node {is: 1}; \\
            \node {Plato: 1}; \\
            \node {\phantom{empty}}; \\
        };
        \node[TextBox, below=of qtf]{Query}; 
        
        \matrix[right=2em of qtf, TFBox] (dtf) {
            \node {Plato: 1}; \\ 
            \node {was: 1}; \\
            \node {Greek: 1}; \\
            \node[DCan] {philosopher: 1}; \\
        };
        \node[TextBox, below=of dtf]{Document}; 

        \node[above=2em of qtf] (dummy_qtf) {};
        \matrix[right=of dummy_qtf, TFBox] (match) {
            \node {who: 1}; \\ 
            \node {is: 0 \phantom{aaaa}}; \\
            \node (node_rho) {Plato: $\rho$}; \\
        }; 
        \draw[->] (qtf) -| (match) ;
        \draw[->] (dtf) -| ([xshift=4pt]match.south) ;
        \node[TextBox, above=of match]{\textit{Match}};

        \matrix[right=1em of dtf, anchor=west, row sep=1mm] (left_tower) {
            \node[Prediction] (z1){$\rho$}; \\ [1mm]
            \node[Predictor] (ls1) {PaRM}; \\
            \node(q1)[QToken]  {who}; 
            \\
        };
        \node(d1)[DToken, fill=blue!10, right=of q1] {philosopher};
        \node[above=15mm of d1] (zs) {};
        \draw[<-] (z1.south) -| ++(0, -0.20);
        \draw[->] (z1.north) |- (node_rho.east);
        \node[right=2em of match, draw, yshift=1em, minimum height=1cm] (bm25) {BM25}; 
        \draw[->] (match.east) |- (bm25);
        \node[Prediction, right=1em of bm25] (y){$y$}; 
        \draw[->] (bm25) -- (y); 
    \end{tikzpicture}
    \caption{The second phase of training PaRM. BM25 computes a relevance score based on the frequency of each query term in the document (\textit{Match}). If a query term (e.g., who) does not appear in the document, the PaRM score $\rho$ for the most relevant document term is used as a discounted query term frequency.  }
    \label{fig:parm_stage2}
\end{figure}

%% file: 4_experiments.tex
\section{Experiments}
\subsection{Implementation}

As a target ranker to be explained,
we use a publicly available cross-encoder, which is fine-tuned from distilled-BERT\footnote{cross-encoder/ms-marco-MiniLM-L-6-v2 from https://huggingface.co/cross-encoder/ms-marco-MiniLM-L-12-v2}. The predictions of this model are used as teacher scores in \autoref{eq:mmse}. We initialized PaRM\ with pre-trained BERT-based-uncased.
The maximum sequence length of the input in the first and second phases of training  PaRM\ is set to 256 and 16 tokens, respectively.

The models are trained on the widely used MS MARCO passage ranking dataset~\cite{msmarco}. 
 BM25 and BM25T use the tokenizer from the Lucene library~\footnote{https://lucene.apache.org/} with the Krovetz stemmer~\cite{krovetz1993viewing}, preferred over the Porter stemmer~\cite{porter1980algorithm} for producing actual words. 

\textbf{Relevance thesaurus construction} PaRM\ scores the candidate term pairs to build the final relevance thesaurus as a global explanation of the full cross-attention ranking model. 
The candidates are drawn from the frequent terms in the MS MARCO corpus. The top 10K frequent terms were considered as query terms, and the top 100K terms as document terms,  resulting in $10^9$ pairs.
Inputs to the PARM model are at the  term level, thus their scores are computed much faster than those of the full cross-encoder that gets long sequences of entire query-document pairs.  
Only candidate term pairs with scores above 0.1 are included in the relevance thesaurus, resulting in a total of 553,864 term pairs. 

\subsection{Evaluations}
\label{sec:evaluation}

We evaluate the BM25T model by exploiting our built relevance thesaurus in two ways: ranking effectiveness and fidelity. 
Ranking effectiveness is measured by standard ranking evaluation metrics that use ground truth judgments. It demonstrates to what extent BM25T can be used for relevance ranking. 
Fidelity expresses the extent to which the BM25T faithfully explains the behavior of the target ranking model, i.e., the cross-encoder model. 

To demonstrate the generalizability of the relevance thesaurus obtained from PaRM, we developed QLT (Query Likelihood with Thesaurus), a variant of BM25T (PaRM) based on the query-likelihood (QL) framework~\cite{ponte1998language}. QLT incorporates the translation language model~\cite{berger1999information} and uses translation probabilities extracted from our relevance thesaurus.
Unlike BM25T which computes the score of a query term  based on the most relevant document term, QLT computes the score by summing the relevance scores of the document terms. 

To provide a baseline comparison and demonstrate the effectiveness of PaRM in building a relevance thesaurus, we re-purposed a local explanation method~\cite{llordes2023explain} as a global explanation~\cite{lundberg2020local}, denoted as BM25T (L to G).
Given a query and ranked candidates documents for it, this explanation method identifies which terms in the document are relevant to the query. 

We adapted it by aligning each document term to the most relevant query term using a cross-encoder and aggregating alignments across 400K training queries.
The aggregated scores of term pairs create a relevance thesaurus, which augments the BM25 scoring function as in BM25T (PaRM) (See \autoref{sec:baseline} for details).

\textbf{In-domain ranking effectiveness.} First, we evaluate ranking effectiveness on three datasets derived from MS MARCO.
TREC DL 2019 and 2020~\cite{trec19dl, trecdl20} contain 43 and 53 queries, respectively. Top-ranked documents are thoroughly judged by NIST assessors, which make them more reliable for evaluating the ranking effectiveness. 

We also used a larger dataset called MS MARCO-dev, which we built by sampling 1,000 queries from the development split of MS MARCO. As this dataset is sparsely judged, with most queries having only one relevant document, we evaluated it using mean reciprocal rank (MRR). MS MARCO-dev will also be used for evaluating fidelity, where more data points are preferable.

\input{table/ndcg}
\autoref{tab:trec_ndcg} shows the ranking effectiveness of methods on the MS MARCO  datasets. 
BM25T with our proposed PaRM shows significant gains ($p<0.01$) over BM25 in all datasets. The obtained gains demonstrate that the distilled relevance thesaurus effectively improves the vocabulary mismatch problem of BM25. In contrast, the BM25T (L to G) does not show consistent improvements. 
BM25T still has a gap from the cross-encoder model, showing room for improvement in future work. Note that we do not include other retrieval models to compare with their ranking effectiveness, as they cannot be used to make a global explanation.
QLT (PaRM) also has better effectiveness compared to QL. Considering that the thesaurus is only tuned for BM25 but not for QL, this result demonstrates the generalization ability of the acquired relevance thesaurus.    

\textbf{Out-of-domain ranking effectiveness.}
The BEIR benchmark~\cite{beir} is a collection of IR datasets and is widely used to measure the generalizability of models without domain-specific training. 
We evaluate the zero-shot ranking effectiveness of the BM25T model over this benchmark, using the same relevance thesaurus distilled from the cross-encoder model that is trained on MS MARCO. 

\autoref{tab:beir_exp} shows evaluation results on the BEIR datasets. Out of the 12 datasets, the performance difference between BM25 and BM25T is statistically significant ($p<0.05$) in 8 datasets. Among these datasets, BM25T outperforms BM25 on 7 datasets, showing a performance closer to that of the cross-encoder. Thus, we conclude that the relevance thesaurus is not limited to the corpus on which it is trained and can effectively perform semantic matching in out-of-domain datasets.

\input{table/beir}

\textbf{Fidelity.}
As our task is a ranking task, we measure faithfulness in terms of the correlation between the scores from the explanations and the target model. 

Given a query $q$ and its corresponding candidate documents $\{d_1, d_2, ..., d_n\}$, the fidelity of an explanation is computed as the correlation between the scores $\{S_b(q, d_1), S_b(q, d_2), ..., S_b(q, d_n)\}$ from the targeted neural model and the global predictor from explanations (e.g., BM25T) $\{S_e(q, d_1), S_e(q, d_2), ..., S_e(q, d_n)\}$. 

For the fidelity score for a dataset, we calculate this correlation for each query and then average them across all queries in the dataset. 
For each query, the top 1,000 documents retrieved by BM25 are used as candidate documents. The Pearson correlation coefficient, which ranges from -1 to 1, is used as a measure of correlation, with 1 indicating the strongest positive correlation.
\autoref{sec:appendix_full_exp} reports results for other correlation measures,  which show similar findings.

In addition to the cross-encoder model, which was used for training PaRM, we apply fidelity evaluation to the other IR models that fine-tune Transformer-based models on MS MARCO. 
Four popular retrieval models are included.
The first two models are TAS-B~\cite{hofstatter2021efficiently} and Splade v2~\cite{spladev2} which are trained using knowledge distillation from the cross-encoders.
The next two are Contriever and Contriever+M~\cite{izacard2021unsupervised}. Contriever is trained with unsupervised learning unsupervised, and Contriever+M is the model that further fine-tunes Contriever using MS MARCO.
 These four models are dual-encoders, where the query and document are independently encoded into vectors using Transformer encoders. 
 
\input{table/corr}

\autoref{tab:corr} shows the fidelity of BM25 and BM25T to these neural retrieval models on MS MARCO-dev.
First, we can observe that in all ranking models, BM25T has higher fidelity than BM25.
Also, the gain is larger on models that are trained on MS MARCO.
Contriever is not trained based on the MS MARCO dataset, on which BM25T showed the lowest fidelity and smallest fidelity gain.

We conclude that the relevance thesaurus can serve to explain the behavior of the models that are trained with similar training data. 

The fidelity evaluations on BEIR datasets also confirm that BM25T is more faithful than BM25 in explaining the cross-encoder model (\autoref{tab:beir_corr}). The fidelity of BM25T is higher than BM25 across all datasets, except in the Quora dataset. The average fidelity across the datasets improved from 0.507 with BM25 to 0.630 with BM25T.

The high fidelity of BM25T to the cross-encoder model is further evidenced by its performance across the BEIR datasets, as shown in \autoref{tab:beir_exp}. In fact, BM25T mirrors the CE's performance drops in the ArguAna and Touche-2020 datasets. This consistency suggests that the relevance thesaurus effectively captures the semantic matching patterns of the CE, even when those patterns lead to decreased performance. Further analysis of the relevance thesaurus could provide insights into why the additional semantic matches sometimes result in worse performance in these specific datasets

\input{4-2_analysis}

%% file: table/ndcg.tex
\newcommand{\nooo}{\phantom{$^{\dag}$}}
\newcommand{\sigt}{$^{\ddagger}$}

\begin{table}[]
    \centering
    \small
    \begin{tabular}{l|c|c|c}
    \toprule 
    Model        & \small{TREC DL19} & \small{TREC DL20} & Dev   \\ 
                     & \small{NDCG@10}    & \small{NDCG@10}    & \small{MRR}   \\ \hline
    BM25           & 0.516\nooo      & 0.503\nooo      & 0.160\nooo \\
    BM25T (L to G) & 0.518\nooo      & 0.501\nooo      & 0.158\nooo \\ 
    BM25T (PaRM)   & 0.550\sigt      & 0.546\sigt      & 0.180\sigt \\ \hline
    QL             & 0.495\nooo      & 0.509\nooo      & 0.153\nooo \\
    QLT (PaRM)     & 0.543\sigt      & 0.540\sigt      & 0.170\sigt \\ \hline
    Cross-encoder   & 0.763\nooo      & 0.739\nooo      & 0.375\nooo \\ 
    \bottomrule
    \end{tabular}
    \caption{Ranking performance on the MS MARCO driven datasets.
    \sigt\ marks the statistically significant difference ($p<0.01$) to the baseline in each group. 
    }
    \label{tab:trec_ndcg}
\end{table}
	

%% file: table/beir.tex
\newcommand{\noo}{\phantom{$^{\dag}$}}
\newcommand{\sig}{$^{\dag}$}

\begin{table}[]
    \small
    \centering
    \begin{tabular}{l|c|c|c}
    \toprule
    Dataset      & BM25  & BM25T & \begin{tabular}[c]{@{}c@{}}Cross\\ Encoder\end{tabular} \\ \hline
    HotpotQA    & 0.633\noo  & 0.641\sig   & 0.725                                                   \\
    DBPedia     & 0.325\noo  & 0.350\sig   & 0.447                                                   \\
    NQ          & 0.307\noo  & 0.332\sig   & 0.462                                                   \\
    Touché-2020 & 0.499\sig  & 0.337\noo  & 0.272                                                   \\
    SCIDOCS     & 0.150\noo  & 0.148\noo  & 0.163                                                   \\
    TREC-COVID  & 0.583\noo  & 0.602\noo  & 0.733                                                   \\
    FiQA-2018   & 0.245\noo  & 0.248\noo  & 0.341                                                   \\
    Quora       & 0.775\sig  & 0.738\noo  & 0.823                                                   \\
    ArguAna     & 0.407\sig  & 0.359\noo  & 0.311                                                   \\
    SciFact     & 0.678\noo  & 0.678\noo  & 0.688                                                   \\
    NFCorpus    & 0.319\noo  & 0.348\dag  & 0.369                                                   \\
    ViHealthQA  & 0.217\sig  & 0.173\noo  & 0.168                                                   \\ \bottomrule
    \end{tabular}
    \caption{The ranking effectiveness measure (NDCG@10) of the methods on BEIR datasets. \sig\ marks the statistically significant difference ($p<0.05$) between BM25 and BM25T
    }
    \label{tab:beir_exp}
\end{table}

%% file: table/corr.tex
\begin{table}[ht]
\centering
    \small 
    \begin{tabular}{l|c|c|c}
    \toprule  
                          &  \multicolumn{2}{c|}{Fidelity}  & Ranking       \\ \hline
    Ranking Model         & BM25    & BM25T  & MRR            \\ \hline
    Cross-encoder         & 0.484   & 0.580  & 0.375           \\
    Splade v2             & 0.490   & 0.583  & 0.335           \\
    TAS-B                 & 0.421   & 0.513  & 0.318           \\
    Contriever            & 0.417   & 0.454  & 0.174           \\
    Contriever+M        & 0.411   & 0.495  & 0.307           \\ 
    \bottomrule
    \end{tabular}
    \caption{Fidelity of the explanations to the ranking models, measured by Pearson correlations on the MS MARCO Dev dataset. Both BM25 and BM25T are considered explanations for the corresponding ranking models. The ranking performance, measured by Mean Reciprocal Rank (MRR), is provided as a reference.}
    \label{tab:corr}
\end{table}

%% file: 4-2_analysis.tex
\subsection{Insights from the relevance thesaurus}
\label{sec:insights}

The relevance thesaurus contains both reasonable and unexpected term pairs, as illustrated in \autoref{ex:table}. Through the analysis of the thesaurus, we identified three interesting findings.

\textbf{Car-brand bias} The thesaurus reveals that the models associate ``car'' with many brand names, but assign higher scores to certain brands over others (\autoref{tab:brand_score_chart}). For example, the pair (``car'', ``Ford'') has a score of 0.39, while (``car'', ``Honda'') has a score of 0.28.

This bias can impact the ranking of documents in the following way: imagine a query containing the term ``car'' and two documents that are identical except for the mentioned car brand - one document includes a high-scoring brand, while the other features a low-scoring brand. Due to the higher score assigned by the thesaurus, BM25T will rank the document with the high-scoring brand above the one with the low-scoring brand. This observation suggests that the neural ranking model is likely to exhibit the same bias, prioritizing documents that mention high-scoring brands over those with low-scoring brands, even when the documents' content is otherwise the same.

\input{table/brand_score_chart}
\textbf{When-year bias} 
 The models exhibit a temporal bias, assigning different scores to various years for the query term ``when'', with much lower scores for years around 2015, when the MS MARCO dataset was constructed (\autoref{fig:when_2015}).
We hypothesize this bias exists because the current year is often less informative for ``when'' questions, as more specific temporal information is typically expected. While effective for 2015 data, it could lead to sub-optimal performance for different current years, such as 2024. 
\input{figures/when_year}

Experiments detailed in \autoref{sec:appendix_bias} validate the presence of these behaviors in state-of-the-art relevance models, supporting that the behaviors specified by the relevance thesaurus are well representing the neural ranking models. 

\textbf{Postfix a} Many thesaurus entries consist of cases where the document term is the query term with an additional ``a'' or ``â'' at the end, such as the (``car'', ``vehicleâ''). This is due to encoding errors in the MS MARCO dataset, where the right quotation marks (') were incorrectly decoded as ``â''. 
The issue is compounded by the BERT tokenizer's normalization of ``â'' to ``a''.
For example, the system might erroneously consider ``cud'' (partially digested food in a cow's stomach) relevant to ``CUDA''  (NVIDIA's parallel computing platform).

%% file: table/brand_score_chart.tex
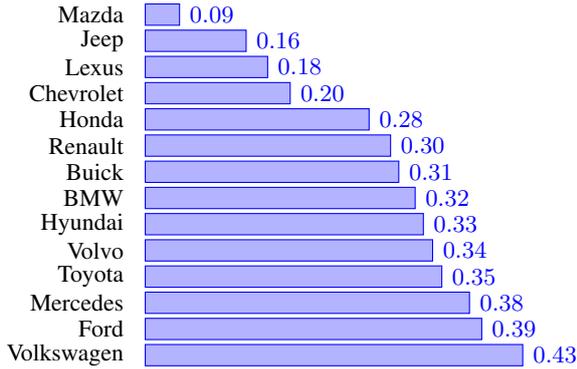
\begin{figure}
    \centering
    \pgfkeys{/pgf/number format/.cd,fixed,fixed zerofill,precision=2}
    \begin{tikzpicture}
    \begin{axis}[
    xbar, 
    width=7cm, 
    height=7cm,
    tick label style={font=\small},
    bar width=8px,
    symbolic y coords={Volkswagen,Ford,Mercedes,Toyota,Volvo,Hyundai,BMW,Buick,Renault,Honda,Chevrolet,Lexus,Jeep,Mazda},
    nodes near coords, 
    nodes near coords align={right},
    nodes near coords style={font=\small},
    ytick distance=1,
    ytick=data,
    xtick={0},
    ytick style={draw=none},
    xtick style={draw=none},
    axis line style={draw=none},
    ]
    \addplot coordinates {
    (0.429,Volkswagen)
    (0.389,Ford)
    (0.377,Mercedes)
    (0.350,Toyota)
    (0.341,Volvo)
    (0.332,Hyundai)
    (0.324,BMW)
    (0.308,Buick)
    (0.300,Renault)
    (0.279,Honda)
    (0.202,Chevrolet)
    (0.180,Lexus)
    (0.159,Jeep)
    (0.094,Mazda)};
    \end{axis}
    \end{tikzpicture}
    \caption{Scores for car brand names against the query term ``car'' based on our relevance thesaurus.}
    \label{tab:brand_score_chart}
\end{figure}

%% file: figures/when_year.tex
\begin{figure}
    \centering    
    \resizebox{7cm}{!}{
    \begin{tikzpicture}
    \pgfkeys{/pgf/number format/.cd,
        fixed,
        precision=999,
        set thousands separator={},
        }
    \begin{axis}[
        xlabel={Year},
        axis y line=left,
        every axis x label/.style={
            anchor=west,
            at={(axis description cs:1,-0.05)}
        },
        ylabel={Rel. Theasaurus},
        xtick={2000, 2010, 2020},
        ytick={0.05, 0.1, 0.15},
        grid=major,
        legend pos=outer north east,
        legend cell align={left},
        legend style={font=\small}
    ]
    
    \addplot[blue] coordinates {
        (2000, 0.15991954505443573) (2001, 0.17009232938289642) (2002, 0.16455566883087158) (2003, 0.19133111834526062)
        (2004, 0.18887905776500702) (2005, 0.18211613595485687) (2006, 0.12253455817699432) (2007, 0.12562014162540436)
        (2008, 0.1550411880016327) (2009, 0.11106028407812119) (2010, 0.1153125986456871) (2011, 0.05630696937441826)
        (2012, 0.039196114987134933) (2013, 0.029429536312818527) (2014, 0.02278118208050728) (2015, 0.004900562111288309)
        (2016, 0.0026514031924307346) (2017, 0.0014096475206315517) (2018, 0.00046073851990513504) (2019, 0.14225299656391144)
        (2020, 0.13050927221775055) (2021, 0.15954212844371796) (2022, 0.030796818435192108) (2023, 0.045765507966279984)
        (2024, 0.06430057436227798)
    }; \label{plot_rt}
    
    \end{axis}
    
    \begin{axis}[
        axis y line=right,
        axis x line=none,
        ylabel={Cross encoder},
        ytick={8, 9},
    ]
    
    \addplot[red] coordinates {
        (2000, 8.402446746826172) (2001, 9.676849365234375) (2002, 9.452437400817871) (2003, 9.660808563232422)
        (2004, 9.34357738494873) (2005, 8.962972640991211) (2006, 8.953495979309082) (2007, 8.873234748840332)
        (2008, 8.64099407196045) (2009, 8.566083908081055) (2010, 8.503437042236328) (2011, 8.311593055725098)
        (2012, 8.33627986907959) (2013, 7.812245845794678) (2014, 7.911380290985107) (2015, 7.778740882873535)
        (2016, 7.777125358581543) (2017, 7.607086181640625) (2018, 7.475591659545898) (2019, 7.903474807739258)
        (2020, 7.827475070953369) (2021, 8.836408615112305) (2022, 8.51842212677002) (2023, 8.793581008911133)
        (2024, 8.792820930480957)
    }; \label{plot_ce}
    \addlegendimage{/pgfplots/refstyle=plot_rt}\addlegendentry{Rel. Thesaurus}
    \addlegendimage{/pgfplots/refstyle=plot_ce}\addlegendentry{Cross encoder}
    \end{axis}
    \end{tikzpicture}
    }
    \caption{Relevance scores for the query term ``when'' and document terms representing years from 2000 to 2024, based on our relevance thesaurus and the cross-encoder model.}
    \label{fig:when_2015}
\end{figure}
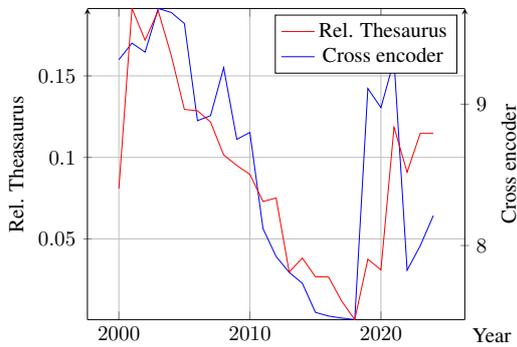

%% file: 5_conclusion.tex
\section{Conclusion}
 
We explored using a relevance thesaurus as a global explanation for neural ranking models. We proposed an effective approach for constructing the thesaurus by training a partial relevance model (PaRM). 
Augmenting the acquired thesaurus into BM25 enhanced its ranking effectiveness and fidelity to the targeted neural ranking model across multiple information retrieval datasets.
Furthermore, the thesaurus uncovered unexpected corpus-specific behaviors and biases of state-of-the-art ranking models, highlighting its value in identifying potential issues and limitations in neural rankers.

We expect a few promising research directions on top of our work. The proposed strategy of using thesaurus to explain a model can be further extended to other Transformer-based models, including generative language models to discover biases on these models. 
For IR applications, effectiveness of BM25T could be improved by considering multiple document terms for each query term and incorporating term location information. These enhancements would better mimic neural ranking models' behavior, potentially leading to more efficient and interpretable sparse retrieval models that more closely match the performance of their neural counterparts. 

%% file: 8_limitations.tex
\newpage
\section*{Limitations}
Our explanation methods have shown the existence of bias in models trained on the MS MARCO dataset. However, the presence of bias does not necessarily indicate inappropriate behavior. In some cases, bias may actually contribute to effective ranking, as certain keywords have a higher likelihood of being relevant due to their ability to be used in different contexts, such as referring to a person or other entities.

While our term replacement experiments were designed to control for context by maintaining identical conditions, real-world documents often exhibit diverse contexts that could potentially diminish the impact of biases. In practice, the contextual differences between documents may result in greater variations in relevance scores compared to the variations caused by biases alone. Consequently, the biases observed in our controlled experiments may have a less significant effect on the ranking of real documents, as the influence of context differences could be more dominant.

The biases identified from the specific relevance thesaurus in our experiments are limited to the ranking models trained on the MS MARCO dataset by fine-tuning BERT-based models. This work has not covered ablations to determine if these biases originated from MS MARCO training data or from language model pre-training of BERT. 

While our experiments demonstrated that training PaRM is effective with MS MARCO data, this approach may not be equally effective in low-resource settings. Specifically, the proposed distillation steps require well-representative queries for the datasets. 

%% file: 9_0_appendix.tex
\appendix

\input{9_1_bias}
\input{9_2_experiments_detail}
\input{9_3_parm_train_detail}

\section{Responsibility statement}

\subsection{Artifact - MS MARCO Dataset}

The MS MARCO dataset~\cite{msmarco}, used as the artifact in this paper, has been carefully curated and anonymized by its creators to protect user privacy and prevent the inclusion of personally identifying information or offensive content. The dataset consists of anonymized search queries and corresponding relevant passages from web pages, processed to remove any personal information.

The MS MARCO dataset is a large-scale information retrieval dataset covering a wide range of domains and topics in the English language. It includes real-world search queries from Bing and corresponding relevant passages from web pages. The dataset is divided into training, development, and testing sets~\footnote{https://microsoft.github.io/msmarco/Datasets}, each containing a substantial number of query-passage pairs. While demographic information is not explicitly provided due to privacy concerns, the dataset is considered representative of diverse information needs and user intents in web search scenarios.

\subsection{AI Assitance}
We acknowledge the use of AI assistants, Claude by Anthropic~\footnote{https://www.anthropic.com/claude} and GPT-4 by OpenAI~\footnote{https://chat.openai.com/}, in the writing process of this paper. These AI assistants provided support in drafting and refining the contents of the paper. However, all final decisions regarding the content, structure, and claims were made by the human authors, who carefully reviewed and edited the generated content.

%% file: 9_1_bias.tex
\section{Unexpected behaviors}
\label{sec:appendix_bias}

In \autoref{sec:insights}, we discovered three unexpected behaviors from the relevance thesaurus. 
This part describes the experiments that support the existence of the behaviors found in the state-of-the-art relevance models. The three behaviors investigated are (1) Postfix a, where the models treat ``a'' at the end of a word as a quotation mark or apostrophe due to encoding errors in the MS MARCO dataset; (2) Car-brand bias, where the models assign higher scores to certain car brand names over others; and (3) When-year bias, where the models exhibit temporal bias in assigning relevance scores to different years for queries containing the term ``when''.

\subsection{Car-brand bias}
\label{sec:car_brand_bias}

\input{table/brand_scores}

The relevance thesaurus reveals that the models associate the query term ``car'' with many brand names, such as ``Ford'' and ``Honda'', but consistently assign higher scores to certain brand names over others (\autoref{tab:brand_score}). To verify if this bias is present in the state-of-the-art relevance ranking models, we designed an experiment using the MS MARCO passage collection.

From the training split of the MS MARCO passage collection, we selected queries that include the term ``car'' but exclude any car brand names or content specific to particular brands.  We then selected documents for each of the queries that satisfy the following criteria:

\begin{enumerate}
    \item The document mentions only one brand name.
    \item The document does not contain any brand-specific information when the brand name is removed.
    \item The document is predicted as relevant by the cross-encoder model.
\end{enumerate}
To filter the documents based on the second and third criteria, we employed keyword-based filtering using a list of car models, ChatGPT-based filtering to identify brand-specific information and manual annotations. For the keyword filtering, we built a list of car models and excluded the documents that contained any of the model names. For ChatGPT-based filtering, we masked the brand name mention of the document and prompted, ``Does this document contain any brand-specific information?'', and if the answer was yes, the document was excluded. This process resulted in 382 query-document pairs, with 29 car brand names considered.

For each query-document pair, the brand name mentioned in the document was replaced by each of the 29 brand names in turn. All the resulting combinations were scored by neural ranking models, yielding a score array of 382 $\times$ 29, where each row represents a query-document pair and each column represents a brand name. In other words, the element at position (i, j) in the array represents the score assigned by the neural ranking model to the $i$-th query-document pair when the brand name is replaced by the $j$-th brand name. The element (i, i) represents the original brand that appeared in the document.

To obtain a single score per brand name, we averaged the scores across the 382 query-document pairs. We then measured the correlation between these average scores and the scores from the relevance table derived from the cross-encoder model. If a neural model exhibits the bias suggested by the representation, we expect a corresponding bias in the modified documents.
\input{table/bias}

\autoref{tab:corr_car} shows the correlation values (fidelity) obtained for each of the ranking models. The results demonstrate that the scores from the thesaurus correlate with scores from the neural ranking models, indicating that our relevance thesaurus can be used to identify possible biases of ranking models.

\subsection{When-year}

The models exhibit a temporal bias where different years have different scores for the query term ``when'', with much lower scores assigned to the years around 2015 compared to other years. Most years (e.g., ``2001'') have high relevance scores for the query term ``when'' in the relevance thesaurus, but the score sharply decreased around 2015, the year when the MS MARCO dataset was constructed (\autoref{fig:when_2015}).

To validate if this bias is in neural models, We measured the predicted scores from the neural ranking models with the query being ``when did North Carolina join IFTA'' and the document being  ``{year} North Carolina join IFTA''.  
\autoref{tab:corr_car} shows the correlation between the scores from the ranking model and the relevance thesaurus.  

The results show that the neural ranking models exhibit a similar temporal bias to the relevance thesaurus, where documents mentioning years around 2016 are scored lower in relevance for queries containing the term ``when''. This correlation confirms that our relevance thesaurus faithfully captures the biases underlying the neural models.

\subsection{Postfix A}
Many entries in the thesaurus consist of cases where the document term is the query term with an additional ``a'' or ``â'' at the end, such as (``car'', ``vehicleâ''). This is due to encoding errors in the MS MARCO dataset, where the right quotation marks (') were incorrectly decoded as ``â''. When combined with the BERT tokenizer normalizing ``â'' to ``a'', it could result in incorrect matching, such as considering the term ``cud'' (food in cows' stomach) to be relevant to ``CUDA'' (parallel computing platform).

To test the postfix hypothesis, we selected 500 relevant query-document pairs that contained a common term. We then appended characters from ``a'' to ``z'' to the document occurrence of this common term. For each appended character, we measured the change in the relevance score assigned by the model.

\input{figures/postfix_a}

The result on the cross encoder as illustrated in \autoref{fig:postfix_a} shows that
while all other alphabet characters result in a large score drop when they are appended to the query term occurrence in the document, appending ``a'' actually results in a small increase of the relevance score. The difference between score changes of ``a'' and other cases are all statistically significant at $p<0.01$.
This supports the existence of many entries in the relevance thesaurus where the document term has the additional character ``a'' at the end of the terms. 

The results, illustrated in Figure \ref{fig:postfix_a}, show that appending ``a'' leads to a small increase in the relevance score, while all other characters result in a significant score drop ($p<0.01$). This supports the existence of entries in the relevance thesaurus where the document term has an additional ``a'' at the end. Similar behaviors were observed on Splade and Contriever-MS MARCO, while Contriever and TAS-B do not exhibit such behavior (\autoref{tab:postfix_a_full}).  

While the existence of encoding errors is known~\cite{msmarcogarbage}, there has been no systematic analysis of how these errors could affect the ranking models. This analysis shows that our thesaurus explanation can be effectively used to discover that the model is using features that may not generalize to other corpora.

\input{figures/append_test_res}

%% file: table/brand_scores.tex
\begin{table}[h]
\centering
\small
\begin{tabular}{cc|cc}
\toprule
\textbf{Brand name} & \textbf{Scores} & \textbf{Brand name} & \textbf{Scores} \\ \hline
Volkswagen & 0.429 & Buick & 0.308 \\
Ferrari & 0.410 & Cadillac & 0.303 \\
Porsche & 0.405 & Renault & 0.300 \\
Fiat & 0.394 & Honda & 0.279 \\
Chrysler & 0.390 & Audi & 0.269 \\
Ford & 0.389 & Peugeot & 0.269 \\
Mercedes & 0.377 & Pontiac & 0.259 \\
Packard & 0.366 & Daimler & 0.219 \\
Oldsmobile & 0.365 & Mitsubishi & 0.212 \\
Toyota & 0.350 & Nissan & 0.205 \\
Jaguar & 0.348 & Chevrolet & 0.202 \\
Volvo & 0.341 & Lexus & 0.180 \\
Hyundai & 0.332 & Jeep & 0.159 \\
BMW & 0.324 & Mazda & 0.094 \\
Bentley & 0.322 &  &  \\ \bottomrule
\end{tabular}
\caption{Scores for each of 29 brand names against the query term ``car'' based on our relevance thesaurus.}
\label{tab:brand_score}
\end{table}

%% file: table/bias.tex
\begin{table}[h]
\centering
    \begin{tabular}{l|c|c}
    \toprule
    Ranking Model  & Car - brand & When - year \\ \hline
    Cross Encoder  & 0.282       & 0.746       \\
    Splade v2      & 0.413       & 0.224       \\
    TAS-B          & 0.367       & 0.484       \\
    Contriever     & 0.419       & 0.422       \\
    Contriever + M & 0.200       & 0.665      \\
    \bottomrule
    \end{tabular}
    \caption{The fidelity of relevance thesaurus focused on two findings.
    The models predict scores on query-document pairs when a brand name or year mention is replaced with another. }
    \label{tab:corr_car}
\end{table}

%% file: figures/postfix_a.tex
\begin{figure}[t]
    \centering
    \includegraphics[width=0.45\textwidth]{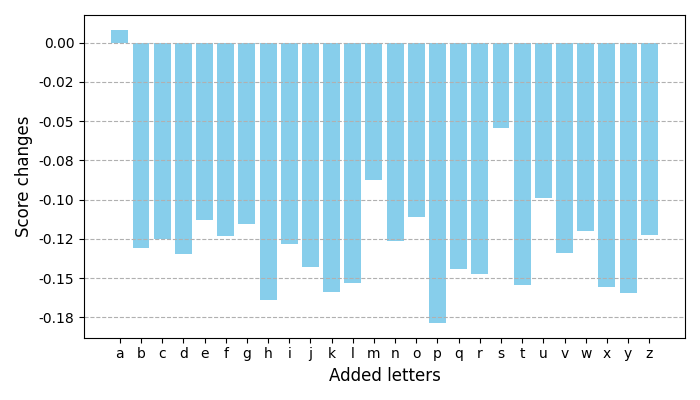}
    \caption{Postfix-a experiment results for the cross encoder. Modifying the matching document term by appending any alphabet results in a large score drop, except `a'.}
    \label{fig:postfix_a}
\end{figure}

%% file: figures/append_test_res.tex
\newpage



\newcommand{\tablefigA}{\includegraphics[width=0.45\textwidth]{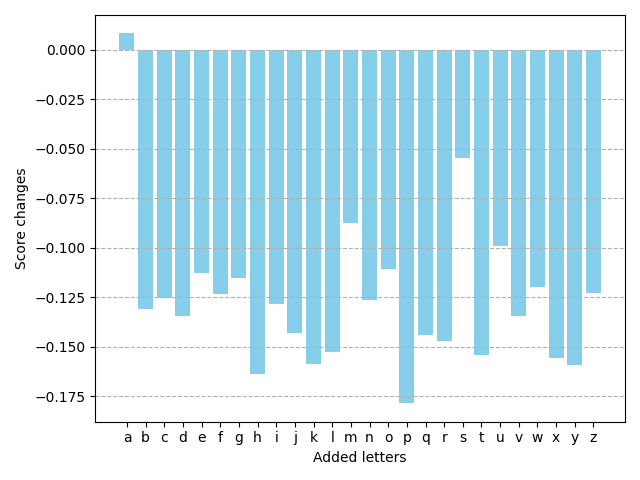}}
\newcommand{\tablefigB}{\includegraphics[width=0.45\textwidth]{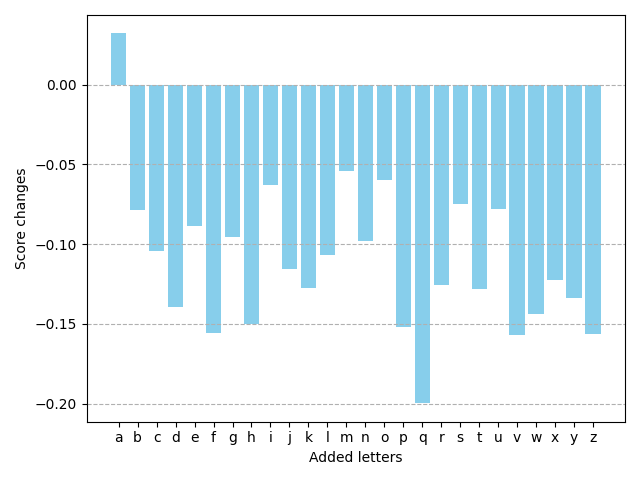}}
\newcommand{\tablefigC}{\includegraphics[width=0.45\textwidth]{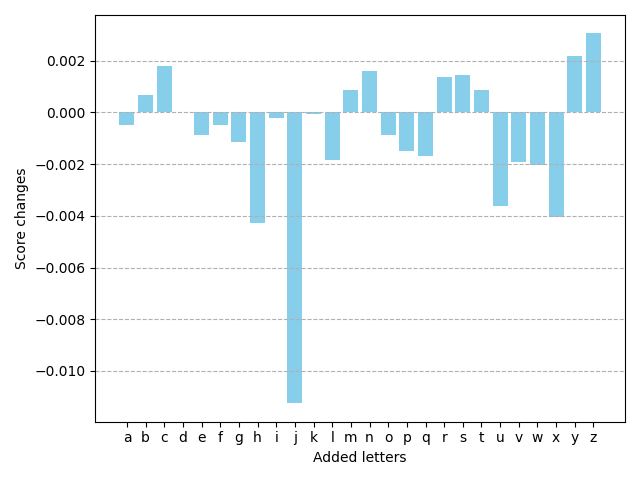}}
\newcommand{\tablefigD}{\includegraphics[width=0.45\textwidth]{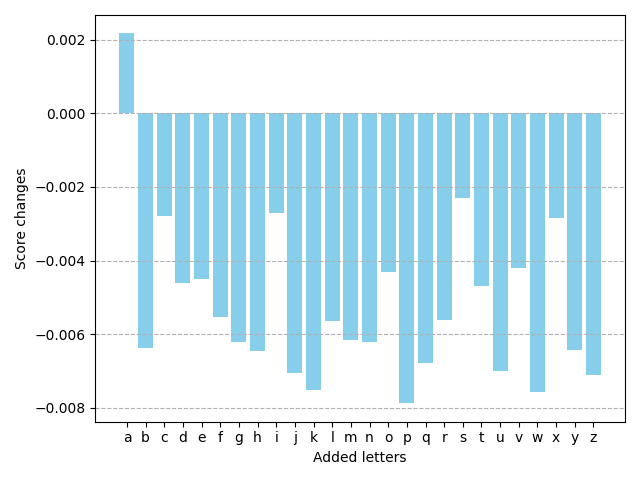}}
\newcommand{\tablefigE}{\includegraphics[width=0.45\textwidth]{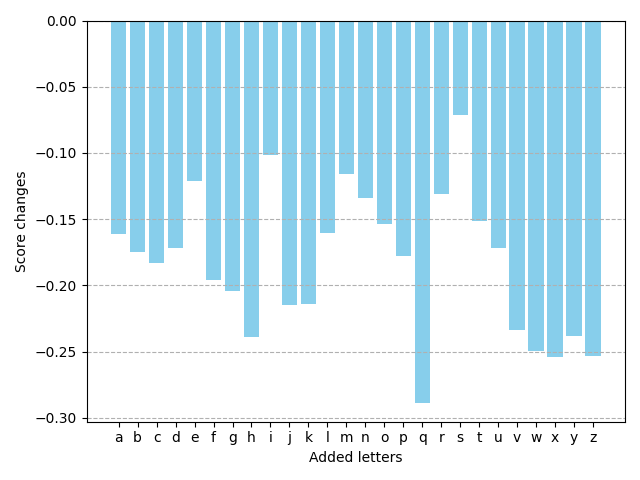}}

\begin{table*}[]
    \centering
    \begin{tabular}{cc}
    \tablefigA & \tablefigB \\
     Cross-encoder    & Splade v2 \\
    \tablefigC & \tablefigD \\
     Contriever    & Contriever+MS MARCO \\
    \tablefigE & \\
     TAS-B    &  \\
    \end{tabular}
    \caption{Postfix-a experiments results. The listed scores are average of $(\text{score after change} - \text{score before change})$, and positive values indicate score increases and negative values indicate score decreases.}
    \label{tab:postfix_a_full}
\end{table*}

%% file: 9_2_experiments_detail.tex
\section{Experiments details}
\label{sec:baseline}

\subsection{BM25T with Local to Global}
As one of our baselines, we adapted the local explanation method proposed by \citet{llordes2023explain} into a global explanation. Their approach provides local explanations for a given query and ranked documents by identifying matching document terms. The local explanation method can be represented as a function:
\[
E: (q, d_1, d_2, ..., d_n)\rightarrow(w_1, w_2, w_3, ... )
\],
where $q$ is a query, $d_i$ is a ranked document, and $w_i$ is a document term that is considered to match the query. 

While it has been argued that local explanations can be converted into global explanations by accumulating them~\cite{janizek2021explaining}, the local explanations for IR models often lack the necessary information to be used globally. Specifically, the document terms ($w_i$) are not attributed to specific query terms, which is a crucial requirement for building a term-level global thesaurus. 

To build a global explanation in the form of a relevance thesaurus, we align each document term to one of the query terms using the following method. Using a cross-encoder ranker, we compute relevance scores between each query term and each document term, considering them as standalone queries and documents, respectively. The query term with the highest score is aligned to the corresponding document term. 

We apply this alignment process to 400K training queries and their re-ranked candidate documents using the cross-encoder. Compared to PaRM, this method uses a similar number of queries but a larger number of documents per query. For each query, 1,000 candidate documents were re-ranked and the top 10 documents were used to select the document terms for explanation building. 

The score between a query term $qt$ and a document term $dt$ is calculated as the number of times $qt$ is aligned to $dt$ divided by the total number of occurrences of $qt$ in the queries. We only included the document terms which appear more than  once, to reduce the noise. \autoref{exp:ltog} shows the result for the different variants for BM25T (L to G).

\begin{table}[htb]
\small
\begin{tabular}{l|c|c}
\toprule 
                       & TREC DL 19 & TREC DL 20 \\ \hline
BM25                   & 0.516      & 0.503      \\ \hline \hline
\multicolumn{3}{l}{BM25T}                        \\ \hline 
PaRM & 0.550      & 0.546      \\
L to G        & 0.504      & 0.501      \\
+ Min frequency filter & 0.518      & 0.501      \\
+ Uniform attribution  & 0.475      & 0.493     \\ \bottomrule
\end{tabular}
\caption{The ranking effectiveness of BM25T (L to G) with different configurations}
\label{exp:ltog}
\end{table}

\subsection{QLT}

We implemented and evaluated the performance of QLT (Query Likelihood with Thesaurus) to test how the relevance thesaurus can be used for different models that the thesaurus was not optimized for. One key difference between QLT and BM25T is that QLT computes the score as the sum of the relevance scores of the document terms, while BM25T computes the query term's score based on the most relevant document term.

In the original query likelihood model, the relevance score for query $q$ and document $d$ is computed as:
\begin{equation}
p(\mathbf{q}|\mathbf{d}) = \prod_i p(q_i |\mathbf{d}),
\end{equation}
where $p(q_i |\mathbf{d})$ is the probability of the query term $q_i$ in the document $d$, calculated as the frequency of $q_i$ in $d$ divided by the length of $d$.

The translation language model for information retrieval~\cite{berger1999information} considers that any document term $w$ can be "translated" into the query term $q_i$ with the translation probability $t(q_i|w)$. The term probability is then computed as:
\begin{equation}
p(q_i |\mathbf{d}) =\sum_w t(q_i|w) p(w|\mathbf{d}),~\label{eq:translation}
\end{equation}
where $t(q_j|w)$ is the translation probability, and $p(w|\mathbf{d})$ is the probability of the word $w$ in the document $d$, again computed as the frequency of $q_i$ in $d$ divided by the length of $d$.
In QLT, we adopt the translation language model while using our relevance thesaurus to compute the translation probability.

\subsection{BM25 and Query Likelihood (QL) configuration}

For the hyper-parameters of BM25, we used the default values ($k_1=0.9$ and $b=0.4$)  as in Pyserini~\cite{Lin_etal_SIGIR2021_Pyserini}.
We used the analyzer configurations as in the Pyserini implementation, which includes normalization, tokenization, stemming, and stopwords removal.

Our query likelihood implementation uses the same tokenizer as in BM25. We used Dirichlet smoothing~\cite{croft2010search}, and tuned the parameter $\mu$ on another validation set, as the default values were ineffective for short passages.     

\section{Full experiments results}
\label{sec:appendix_full_exp}

\subsection{Different fidelity metrics}
Many works have used different metrics for the fidelity evaluation of ranked lists in IR tasks, such as the overlap of \textbf{top-k} in the ranked list~\cite{llordes2023explain}, agreement rates of \textbf{pairwise} preference agreement in the ranked list~\cite{lyu2023listwise}, or \textbf{Kendall }rank correlations~\cite{pandian2024evaluating}.

We chose the Pearson correlation as the main metric because it considers the magnitude of score differences rather than just the rankings. This is particularly important in IR tasks, where the emphasis is on differentiating a few highly relevant documents from the many non-relevant ones.

While IR tasks are indeed ranking tasks, they prioritize top-ranked documents, which are more likely to be relevant. As a result, ranking correlations like the Kendall rank correlation may be less appropriate, as they are more affected by the ranking of non-relevant documents, which outnumber relevant ones. We also consider the overlap rates of top-k to be less desirable for two reasons: first, the number of relevant documents is unknown; second, it does not account for score differences among the top-k items, which are crucial for reliable IR metrics like NDCG.
Nevertheless, we have included results for other fidelity metrics (\autoref{tab:fidelity_full}), which consistently confirm the improvements of BM25T (PaRM) over BM25.

\input{table/fidelity_full}

\subsection{Fidelity on BEIR}
In \autoref{sec:evaluation}, we include only the ranking effectiveness of BM25T on the BEIR dataset, and not the fidelity (correlation).
\autoref{tab:beir_corr} shows the correlation of the scores when BM25 or BM25T is considered as an explanation and compared to scores from the cross-encoder model. In most datasets, the correlations increased, with the only exception of the Quora dataset.  
\input{table/beir_corr}

The improvements in ranking performance are consistent with the increased correlations observed in most datasets.

%% file: table/fidelity_full.tex
\begin{table*}[htb]
\begin{tabular}{c|cc|cc|cc|cc}
\toprule
\multicolumn{1}{l}{} & \multicolumn{2}{|c}{Pearson $r$} & \multicolumn{2}{|c}{Kendall $\tau$} & \multicolumn{2}{|c}{Pairwise} & \multicolumn{2}{|c}{Top-k overlap} \\ \hline
Model                & BM25          & BM25T         & BM25           & BM25T         & BM25          & BM25T        & BM25        & BM25T      \\ \hline
Cross Encoder        & 0.484         & 0.580         & 0.260          & 0.341         & 0.632         & 0.672        & 0.293       & 0.334      \\
Splade v2            & 0.490         & 0.583         & 0.268          & 0.346         & 0.636         & 0.674        & 0.304       & 0.345      \\
TAS-B                & 0.421         & 0.513         & 0.228          & 0.304         & 0.616         & 0.653        & 0.256       & 0.293      \\
Contriever           & 0.417         & 0.454         & 0.230          & 0.259         & 0.617         & 0.631        & 0.273       & 0.289      \\
Contriever + M  & 0.411         & 0.495         & 0.225          & 0.294         & 0.615         & 0.648        & 0.264       & 0.303      \\ \bottomrule
\end{tabular}
\caption{Fidelity of BM25 and BM25T when measured by different metrics.  }
\label{tab:fidelity_full}
\end{table*}

%% file: table/beir_corr.tex
\begin{table}[]
\centering
\begin{tabular}{c|c|c}
\toprule
Dataset     & BM25  & BM25T \\ \hline
HotpotQA    & 0.535 & 0.647 \\
DBPedia     & 0.477 & 0.612 \\
NQ          & 0.474 & 0.658 \\
Touché-2020 & 0.403 & 0.689 \\
SCIDOCS     & 0.598 & 0.663 \\
TREC-COVID  & 0.276 & 0.705 \\
FiQA-2018   & 0.481 & 0.514 \\
Quora       & 0.659 & 0.640 \\
ArguAna     & 0.656 & 0.722 \\
SciFact     & 0.634 & 0.677 \\
NFCorpus    & 0.584 & 0.626 \\
ViHealthQA  & 0.314 & 0.410 \\
\bottomrule
\end{tabular}
    \caption{Fidelity (Pearson correlation) of the BM25 and BM25T as explanation to the cross encoder ranking model. }
    \label{tab:beir_corr}
\end{table}

%% file: 9_3_parm_train_detail.tex
\section{PaRM implementation details}
\label{sec:appendix_sample}

In the first phase, 
PaRM calculates a relevance score for the given query-document by generating scores for two inputs, ($q_1, d_1$) and ($q_2, d_2)$, which are built from the given query $q$ and document $d$.
Each of the two inputs is then scored through PaRM and these two scores are summed as the relevance score for the query-document pair. Using the relevance label for the query and document, PaRM is trained end-to-end to predict relevance for partial sequences of the query and document without fine-grained labels. 

We build $q_1$ by extracting a continuous span from $q$ and build $q_2$ with the remaining tokens, leaving a [MASK] token where $q_1$ was extracted.  
Given $q_1$, $q_2$, and $d$, we build $d_1$ and $d_2$ by masking some tokens of the document $d$, while keeping tokens that are likely to be relevant to the corresponding $q_i$. Both $d_1$ and $d_2$ can be composed of many non-continuous spans.

\subsection{Building partial segments for the first phase}
The tokens to be deleted are selected so that the deleted tokens in $d_i$ are less likely to be important for the corresponding query partition $q_i$. To estimate which tokens of the document are less likely to be important for a query partition, we use the attention scores from a canonical cross-encoder model which takes the concatenation of whole query $q$ and document $d$ as an input. 

The scoring is done in the following steps.
\begin{enumerate}
    \item We collect normalized attention probabilities from all the layers and heads of the Transformer network.
    As a result, we get a four dimension tensor of $W \in \mathbb{R}^{L \times L \times M \times H}$, where $L$ is the sequence length, $M$ is the number of layers, and $H$ is the number of attention heads in each layer. 
        $W_{ijlk}$ denote the attention probability for the $i$-th token to attend to the $j$-th token in the $k$-th attention head of the $l$-th layer.
    \item We average $W$ over the last two dimensions, which correspond to different layers and heads, and get a two-dimensional matrix $A$.
    \begin{equation}
    A_{ij} = \underset{l}{\sum} \underset{k}{\sum} W_{ijlk}
    \end{equation}
    
    \item Let $|q|$  be the number of tokens in the query and $|d|$ to be the number of tokens in the document. When a [CLS] token and [SEP] tokens are combined with the query and document tokens, the query tokens are located from the second token to $(|q|+1)$-th token, and the document tokens are located from $(|q|+3)$-th token to $(|q|+|d|+2)$-th.
    
    Then, $A_{2:|q|+1, |q|+3:|q|+|d|+2}$ indicates the averaged probability that query tokens to attend to document tokens, and $A_{ |q|+3:|q|+|d|+2, 2:|q|+1}$ indicates averaged probability that document tokens attend to query tokens.  
    \item By transposing the latter matrix and adding it to the first, we obtain $S$. In this resultant matrix, $S_{ij}$ indicates the degree of attention between the $i$-th token of the query and $j$-th token of the document. 
\end{enumerate}

To avoid splitting a word into subwords, the token selection was performed at the word-token level instead of the subword-token.  
The subword-token-level scores are converted by taking maximum scores.

We select the tokens with the lowest attention probability for the corresponding query partition as ones to be deleted. 

The number of tokens to be deleted from the document is randomly sampled so that it can have variable inputs starting from a single token to nearly a full sequence. 
The number of deleted tokens $ m $ is sampled from a normal distribution with the mean and standard deviation being half of the document length $ |d| $. The sampled number is capped at a minimum of 1 and a maximum of $|d|-1$.

\subsection{Second Phase}

In the second phase, PaRM is trained on word pairs, where the query term does not appear in the document. Here are a few clarifications about the details.

To select term-pair candidates, we pre-computed scores for the term pairs by limiting them to the frequent terms. Similar to the relevance thesaurus itself, the top 10K frequent terms were considered as query terms, and the top 100K terms as document terms. The terms without scores are not selected for training.  

The terms that are fed to PaRm are stemmed in the same way they are used for BM25T. However, stopwords are NOT excluded for this step, while BM25T excludes them. 

\subsection{Training configurations}

Both phases use the query-documents triplets provided with MS MARCO passage ranking dataset~\cite{msmarco}, which makes about 400,000 training instances. 

For the first stage training of \modelname , we apply early stopping based on the loss validation set. We used the batch size of 16 and learning rate of 2e-5, which were not tuned. 

For the second stage of training of \modelname , we tuned the learning rate and batch sizes based on its loss on holdout split and BM25T augmented performance (MRR) in a validation set that is separately sampled from MS MARCO dev.
The reported model used a batch size of 256 and a learning rate of 1e-5.

\subsection{Computational cost}
We used four NVIDIA GeForce GTX 1080 Ti GPUs for training. Both the first and second stages of training took less than 10 hours each. The inference for relevance thesaurus construction was run on approximately 100 GPUs, including GeForce GTX 1080 Ti and GTX Titan X models, and took about 700 GPU hours.